\documentclass[a4paper, 12pt, english]{article}

\usepackage[utf8]{inputenc}
\usepackage{amsmath,amssymb}
\usepackage{graphicx}
\usepackage{subfig}
\usepackage[colorinlistoftodos]{todonotes}

\usepackage{amsmath}
\usepackage{amssymb}
\usepackage{graphicx}
\usepackage{algorithm2e}
\usepackage{breqn}
\usepackage{multirow}
\usepackage{capt-of}
\usepackage[T1]{fontenc}
\usepackage{letltxmacro}
\usepackage{listings}
\usepackage{color}

\definecolor{dkgreen}{rgb}{0,0.6,0}
\definecolor{gray}{rgb}{0.5,0.5,0.5}
\definecolor{mauve}{rgb}{0.58,0,0.82}

\usepackage{indentfirst}
\usepackage{verbatim}
\usepackage{textcomp}
\usepackage{gensymb}

\usepackage{relsize}

\usepackage{lipsum}
\usepackage{xcolor}
\usepackage{xparse}
\NewDocumentCommand{\myrule}{O{1pt} O{2pt} O{black}}{%
  \par\nobreak 
  \kern\the\prevdepth 
  \kern#2 
  {\color{#3}\hrule height #1 width\hsize} 
  \kern#2 
  \nointerlineskip 
}
\usepackage[section]{placeins}

\usepackage{booktabs}
\usepackage{colortbl}%

\usepackage[obeyspaces]{url}
\usepackage{etoolbox}
\usepackage[colorlinks,citecolor=black,urlcolor=blue,bookmarks=false,hypertexnames=true]{hyperref} 

\usepackage{geometry}
\begin{document}

\begin{titlepage}
\begin{center}

\includegraphics[width=0.5\textwidth]{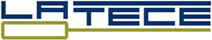}\\
\textbf{\large Laboratoire de Recherches sur les Technologies du Commerce Électronique}\\[0.2cm]

\vspace{10pt}

\includegraphics[width=0.5\textwidth]{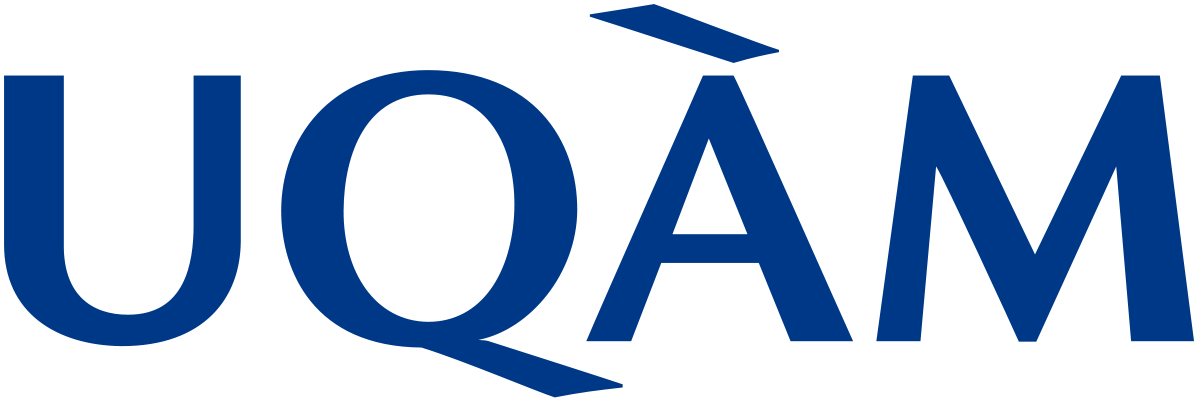}\\[1cm]

\textbf{\LARGE Universit\'e du Qu\'ebec \`a Montr\'eal}\\[0.5cm] 
\vspace{20pt}

\par
\vspace{15pt}
\myrule[1pt][7pt]
\textbf{\Large Identifying KDM Model of JSP Pages}\\
\myrule[1pt][7pt]

\vspace{35pt}
\large{\textbf{Anas Shatnawi, Hafedh Mili, Manel Abdellatif, Ghizlane El Boussaidi, Jean Privat, Yann-Ga\"el Gu\'eh\'eneuc, Naouel Moha}}\\[0.3cm]

\textit{LATECE Technical Report 2017-3, LATECE Laboratoire, Universit\'e du Qu\'ebec \`a Montr\'eal, Canada}

\vspace{200pt}
\small{July, 2017}

\end{center}

\par
\vfill
\begin{center}
\end{center}
\end{titlepage}



\newpage

\begin{center}
\textbf{\Large Identifying KDM Model of JSP Pages}\\[0.9cm]

{Anas Shatnawi\footnote{anasshatnawi@gmail.com}, Hafedh Mili\footnote{mili.hafedh@uqam.ca}, Manel Abdellatif, Ghizlane El Boussaidi, \\Jean Privat, Yann-Ga\"el Gu\'eh\'eneuc, Naouel Moha}\\[0.5cm]

\textit{LATECE Technical Report 2017-3, LATECE Laboratoire, Universit\'e du Qu\'ebec \`a Montr\'eal, Canada}

\end{center}

\begin{abstract}
	In this report, we propose our approach that identifies a KDM model of JSP pages. Our approach is based on two main steps. The first one aims to reduce the problem space by translating JSP pages into Java Servlets where we can use existing tools to identify a KDM model. The second step aims to complete the resulting KDM model by identifying dependencies of JSP tags that are not codified by the translation step. 
\end{abstract}

\section{Introduction}
\label{Introduction}
J2EE applications are implemented based on components developed using a variety of technologies. These technologies are written following Java code (e.g., JavaBeans, Managed Beans, etc.) or scripting languages using XML tags (e.g., JSP, JSF).

To best to our knowledge, there is no approach/tool that identifies a KDM model for the JEE technologies implementing using scripting languages (HTML, JSP, JSF, ASP, etc.). 

Therefore, in this paper, we propose an approach that identifies a KDM model of JSP pages.
Our approach is based on two main steps. The first one aims to reduce the problem space by translating JSP pages into Java Servlets where we can use existing tools to identify a KDM model. To do so, we use \textit{Jasper} tool for translating JSP to Java and the \textit{MoDisco} tool for extracting KDM from Java. The second step aims to complete the resulting KDM model by identifying dependencies of JSP tags that are not codified by the translation process. We build a table of the set of the tags and their related attributes and developed a set of tools that parse JSP pages and configuration files to identify dependencies related to these tags.

The rest of this report is organized as follows. We provide an overview of the proposed approach in Section \ref{sec:overview}. Then, we show how we reduce the problem space by translating JSP pages to Java Servlets in Section \ref{sec:reduction-problem}. Next, we discuss the codification of the JSP tags in Section \ref{sec:codification-remaining-tags}. We finish by concluding the report in Section \ref{sec:conclusion}.

\section{Overview of the Proposed Approach}
\label{sec:overview}
The gaol is to represent the implementation of JSP pages in a KDM model. This includes both the program elements and their related dependencies. The process of the proposed approach is based on two main steps: 

\paragraph{Step 1: The Reduction of the Problem Space}

The main idea behind this step is to map our problem to another already solved problem. Thanks to the MoDisco tool which supports the transformation of the normal Java code to KDM models. Therefore, we decide to translate JSP pages to equivalent Java code. 

\paragraph{Step 2: The Codification of Dependencies Related to JSP Tags}

In the first step, we solve a part of the problem by translating the JSP implementation to Java code. However, there are some JSP tags that are not translated to Java code. Therefore, we aim, in this step, to identify these tags and codify their related dependencies in the KDM model.

\begin{figure*}[!h]
	\begin{center}
		\includegraphics[width=\textwidth]{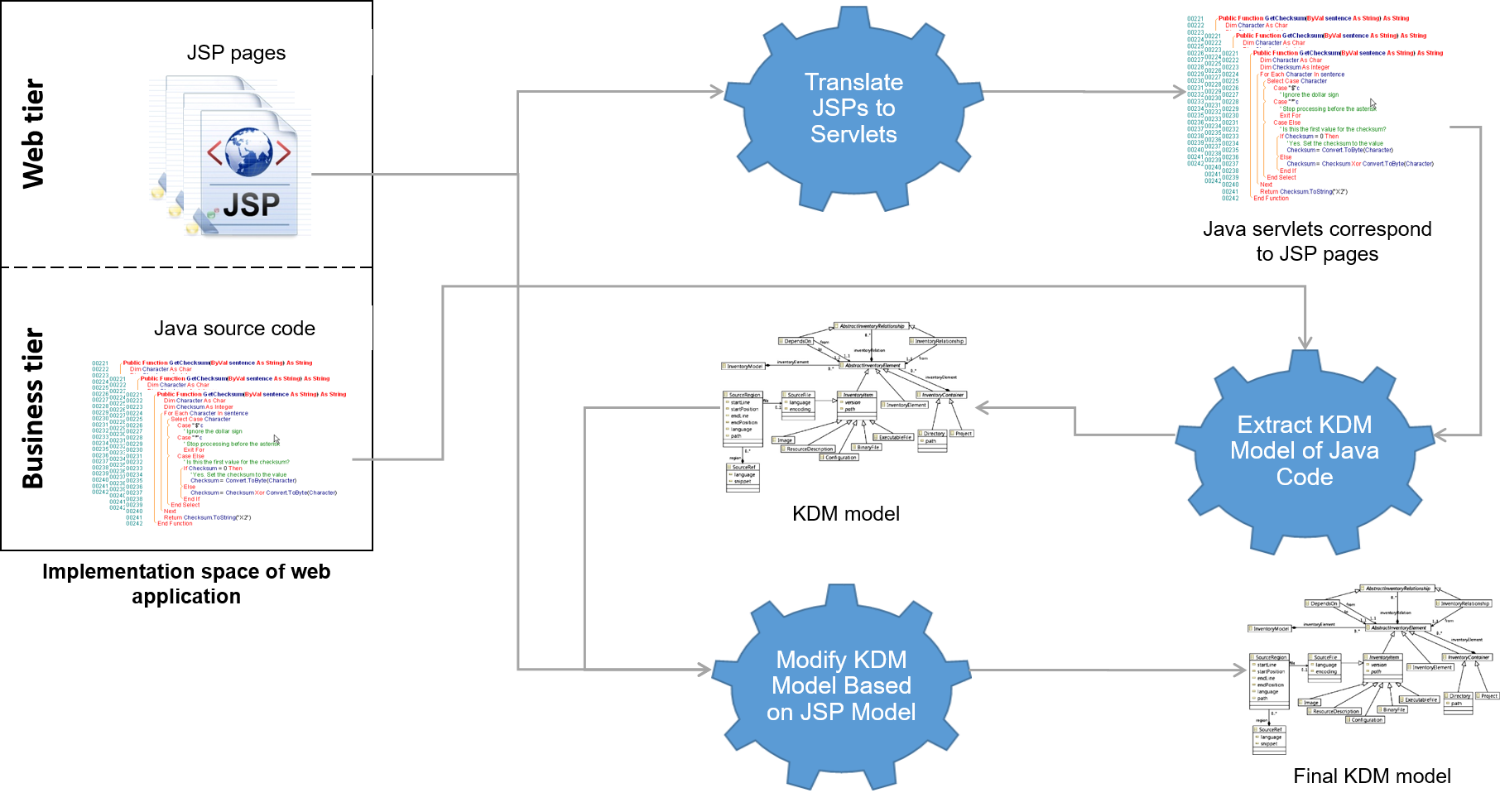}
		\caption{Process of identifying KDM model of JSP pages}
		\label{fig:process}
	\end{center}
\end{figure*}

The main process of our approach is shown in Figure \ref{fig:process}. 

\section{The Reduction of the Problem Space}
\label{sec:reduction-problem}
We select to translate the implementation of JSPs to Servlets (Java code) due to: 

\begin{enumerate}
\item Servlets represent the underlying Java implementation of server pages. Thus, they support the life cycle management of JSP pages,

\item the availability of an open-source tool that translates JSP pages into Servlets such that the resulting Servlets exactly provide the same functionalities (output) of the corresponding JSP pages, and

\item the availability of an open-source tool that identifies KDM models of Servlets.
\end{enumerate}

In this section, we present how to translate JSPs to Servlets using the Jasper tool. Then, we show how to identify a KDM model of these resulting Servlets using the MoDisco tool.

\subsection{The Translation of JSPs to Servlets Using Jasper}

\begin{figure*}[!h]
	\begin{center}
		\includegraphics[width=\textwidth]{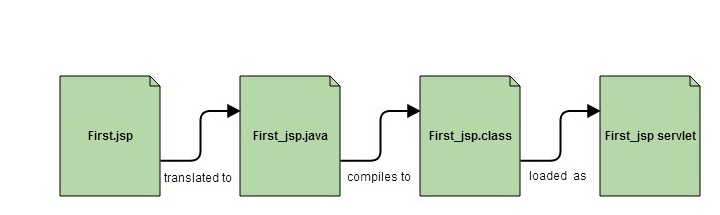}
		\caption{Process of converting JSP pages into Servlets by Jasper}
		\label{fig:jsp-to-servlet-translation}
	\end{center}
\end{figure*}

To translate JSP pages to Servlet classes, we use the \textit{Jasper} tool that is provided as a part of the Apache Tomcat server \cite{apache-tomcat}. The process used by Jasper to translate JSP pages into Servlets is presented in Figure \ref{fig:jsp-to-servlet-translation}.

\begin{figure*}[!h]
	\begin{center}
		\includegraphics[width=0.75\textwidth]{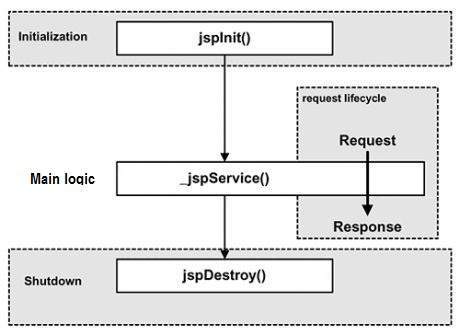}
		\caption{The life cycle of JSP pages}
		\label{fig:jsp-life-cycle}
	\end{center}
\end{figure*}

The Java class of a resulting Servlet is structured in three methods following the JSP life cycle that is shown in Figure \ref{fig:jsp-life-cycle}. These methods are \textit{\_jspInit(...)}, \textit{\_jspService(...)} and \textit{\_jspDestroy(...)} that are respectively used to initialize the Servlet, to serve the requests arrived to the Servlet from clients and to remove the resources.

In the following sub-sections, we present the rules that Jasper used to translate a JSP page to a Servlet, the implementation of Jasper based on ant project and an example of of a JSP page and its translated Servlet.

\subsubsection{Transformation Rules of JSPs to Servlets Based on Jasper}
The translation process of each JSP tag is done sequentially as it is implemented in the corresponding JSP page. The JSP tags is translated based on the following rules:

\begin{itemize}
\item \textbf{JSP scriptlet tags} (e.g., \textit{<\% code fragment \%>}) that are used to insert Java code inside the JSP pages are represented in the Servlet class based on the same code fragment that it contains. 

For example, the \textit{scriptlet} \textit{<\% for (int i=0; i<10; i++) \%>} is translated to \textit{for (int i=0; i<10; i++)}.
	
\item \textbf{JSP declaration tags} (e.g., \textit{<\%! declaration; [ declaration; ]+ ... \%>}) that are used for variable declarations are converted into equivalent variable declarations.

For example, this declaration \textit{<\%! int i=0; !\%>} is translated to \textit{int i=0;}.

\item \textbf{The reference to a JavaBeans component} is converted by creating an instance of the corresponding class. The references to setter/getter methods are  realized through normal Java method invocations.

For example, the \textit{<jsp:useBean id="myBeans" class="package.BeansClass" scope="session">} tag is transformed into an object instantiation of \textit{package.BeansClass}, and \textit{<jsp:getProperty name="myBeans" property="firstName"> </jsp:getProperty>} into a method invocation to the \textit{getFirstName()} method of this object instance.

\item \textbf{The use of a custom tag handler} is realized in terms of a set of method invocations related to the life cycle management methods of the tag handler. Such methods are \textit{doStartTag(), doEndTag() setAttribute()} etc.

\item \textbf{Other JSP tags and HTML/XML tags} are only written as string literal parameters attached in the \textit{out} object related to the response object (\textit{out.write("...");}). 

For example,  the \textit{<jsp:include page="/myPage.jsp." flush="true" />} tag is converted as \textit{out.write("<jsp:include page="/myPage.jsp." flush="true" />");}. These tags need further processing (c.f., Section \ref{sec:codification-remaining-tags}).

\end{itemize}

A result of these transformation rules, we confirm that several program elements and their related dependencies of JSP pages are converted to Java code. Based on this Java code, we will identify a KDM model.  
\subsubsection{The Implementation of Jasper Tool Based on Ant Code to Translate JSPs to Servlets in a Web Application}

To be able to convert JSPs into Servlets, the \textit{Jasper} tool requires having the \textit{Java SDK}\footnote{This video demonstrates how to install \textit{Apache Tomcat}\footnote{This video demonstrates how to install the Apache Tomcat: https://www.youtube.com/watch?v=kLgquZ2FiuQ} \cite{apache-tomcat}, the Java SDK: https://www.youtube.com/watch?v=asoDE3AQXBA} \cite{java-sdk} and \textit{Apache Ant}\footnote{This video demonstrates how to install the Apache Ant: https://www.youtube.com/watch?v=39OwnrUlz0k} installed on your machine. 

In \cite{Jasper}, Apache Tomcat provides the ant code that translates JSP pages of a given JEE Web application into Servlets based on Jasper.
Listing \ref{antToCompileJSP} shows this code where one has to determine the directory of Apache Tomcat (i.e., \textit{<property name="tomcat.home" ...}), the directory of the resulting Java code (\textit{<property name="webapp.path"...}) and the directory related to the other output like the new generated \textit{web.xml} and the \textit{.class} compiled files. 

\begin{lstlisting}[caption=Ant code to convert JSP pages into Java servlets, label=antToCompileJSP]

<project name="Webapp Precompilation" default="all" basedir=".">

		<property name="tomcat.home" value="C:\Program Files\Apache Software Foundation\Tomcat 5.5" />
		<property name="webapp.path" value="C:\...ResultDirectory" />
		<property name="javacode.path" value="C:\...JavaCodeDirectory" />
		
		<target name="jspc">
		
				<taskdef classname="org.apache.jasper.JspC" name="jasper2">
				<classpath id="jspc.classpath">
						<pathelement location="${java.home}/../lib/tools.jar" />
						<fileset dir="${tomcat.home}/bin">
								<include name="*.jar" />
						</fileset>
						<fileset dir="${tomcat.home}/server/lib">
								<include name="*.jar" />
						</fileset>
						<fileset dir="${tomcat.home}/common/lib">
								<include name="*.jar" />
						</fileset>
				</classpath>
				</taskdef>
				
				<jasper2 validateXml="false" uriroot="${webapp.path}" webXmlFragment="${webapp.path}/WEB-INF/generated_web.xml" outputDir="${javacode.path}/jsps" />
				
		</target>
		
		<target name="compile">
		
				<mkdir dir="${webapp.path}/WEB-INF/classes" />
				<mkdir dir="${webapp.path}/WEB-INF/lib" />
				
					<javac destdir="${webapp.path}/WEB-INF/classes" optimize="off" debug="on" failonerror="false" srcdir="${javacode.path}/jsps" excludes="**/*.smap">
							<classpath>
									<pathelement location="${webapp.path}/WEB-INF/classes" />
									<fileset dir="${webapp.path}/WEB-INF/lib">
											<include name="*.jar" />
									</fileset>
									<pathelement location="${tomcat.home}/common/classes" />
									<fileset dir="${tomcat.home}/common/lib">
											<include name="*.jar" />
									</fileset>
									<pathelement location="${tomcat.home}/shared/classes" />
									<fileset dir="${tomcat.home}/shared/lib">
											<include name="*.jar" />
									</fileset>
									<fileset dir="${tomcat.home}/bin">
											<include name="*.jar" />
									</fileset>
							</classpath>
							<include name="**" />
							<exclude name="tags/**" />
				</javac>
		
		</target>
		
		<target name="all" depends="jspc,compile">
		</target>
		
		<target name="cleanup">
				<delete>
						<fileset dir="${javacode.path}/jsps" />
						<fileset dir="${webapp.path}/WEB-INF/classes/org/apache/jsp" />
				</delete>
		</target>

</project>
\end{lstlisting}

\subsubsection{An Example of Translating a JSP to a Servlet}

Listing \ref{JSPCodeExample} shows an example of a JSP page that is translated to an equivalent Java Servlet presented in Listing \ref{JSPinServlet}. In this example, the translated code only encapsulates the HTML tags in terms of string literals that will be handled by the Web container. 

\begin{lstlisting}[caption=An example JSP page, label=JSPCodeExample]
<HTML>
<HEAD><TITLE>Powers of 2</TITLE></HEAD>
<BODY>
<CENTER>
<H2>Behold The Powers Of 2</H2>
</CENTER>
<TABLE BORDER="2" ALIGN="center">
<TH>Exponent</TH>
<TH>2^Exponent</TH>
<% for (int i=0; i<10; i++) {%>
<TR>
<TD><%= i%></TD>
<TD><%= Math.pow(2, i)%></TD>
</TR>
<% } //end for loop %>
</TABLE>
</BODY>
</HTML>
\end{lstlisting}

\begin{lstlisting}[caption=The resulting Java Servlet based on the example JSP page, label=JSPinServlet]
package myPakage;
import java.io.*;
import javax.servlet.*;
import javax.servlet.http.*;
public class PowersOf2 extends HttpServlet
{
public void service(HttpServletRequest request,
HttpServletResponse response)
throws IOException, ServletException
{
response.setContentType("text/html");
ServletOutputStream out = response.getOutputStream();
out.print("<HTML>");
out.print("<HEAD><TITLE>Powers of 2</TITLE></HEAD>");
out.print("<BODY>");
out.print("<CENTER>");
out.print("<H2>Behold The Powers Of 2</H2>");
out.print("</CENTER>");
out.print("<TABLE BORDER='2' ALIGN='center'>");
out.print("<TH>Exponent</TH><TH>2^Exponent</TH>");
for (int i=0; i<10; i++)
{
out.print("<TR><TD>" + i + "</TD>");
out.print("<TD>" + Math.pow(2, i) + "</TD>");
out.print("</TR>");
} //end for loop
out.print("</TABLE></BODY></HTML>");
out.close();
}
}

\end{lstlisting}

\subsection{Identification of KDM Model of Java Code Using MoDisco}
In the previous section, we translated JSP pages into Java code based on Servlets. In this section, we show how to identify the KDM model of these translated Servlets. 

The \textit{MoDisco} tool\footnote{This video shows how to install MoDisco: https://www.youtube.com/watch?v=9PAspfzJn2E} provides a set of \textit{APIs} and \textit{Discoverers} that allow to extract KDM models of a given Java project based on the static analysis of its source code. The extracted model contains the complete Abstract Syntax Tree (AST) of all statements in the source code.

MoDisco offers a \textit{Java Discoverer} as an Eclipse plug-in that allows one to easily extract the KDM model. One can follow these steps to get a KDM model.

\begin{enumerate}
\item \textbf{Adding Dependencies of MoDisco Plug-ins:} One has to added dependencies to three plug-ins in his/her project at the \textit{Require-Bundle} in the \textit{Manifest.MF} file\footnote{See this to know how to add dependencies: https://help.eclipse.org/mars/index.jsp?

topic=\%2Forg.eclipse.pde.doc.user\%2Fguide\%2Ftools\%2Feditors\%2Fmanifest_editor\%2Fdependencies.htm}. These are: 

\begin{itemize}
\item \textit{org.eclipse.gmt.modisco.java}

\item \textit{org.eclipse.modisco.java.discoverer}

\item \textit{org.eclipse.modisco.infra.discovery.core}
\end{itemize}

\item \textbf{Selecting the Good Discoverer Based on Your Input:}
MoDisco's APIs offer a set of classes such that each one takes a different input (e.g., project, class etc.). Table \ref{table:modisco-discoverer} shows the list of MoDisco's discoverers and their required input instances. One has to decide what is the discoverer that confirms to his/her input.

\begin{table}[h]
\centering
\caption{List of MoDisco discoverers and their inputs}
\label{table:modisco-discoverer}
\begin{tabular}{|l|l|}
\hline
\multicolumn{1}{|c|}{\textbf{Dicoverer Class}}   & \multicolumn{1}{c|}{\textbf{Required Input}} \\ \hline
DiscoverJavaModelFromJavaProject        & IJavaProject                        \\ \hline
DiscoverJavaModelFromProject            & IProject                            \\ \hline
DiscoverJavaModelFromClassFile          & IClassFile                          \\ \hline
DiscoverJavaModelFromLibrary            & IPackageFragmentRoot                \\ \hline
\end{tabular}
\end{table}

\item \textbf{Implementing the KDM Discoverer in Your Code:} One has to create an instance of the selected discoverer and give it the input. Listing \ref{discoverer-implementations} shows an example of code that implements a MoDisco discoverer from a given Java project.

\end{enumerate}

\begin{lstlisting}[caption=Example of code that implements a MoDisco discoverer, label=discoverer-implementations]
...
DiscoverJavaModelFromJavaProject discoverer = new DiscoverJavaModelFromJavaProject();
javaDiscoverer.discoverElement(javaProject, monitor);
Resource javaResource = javaDiscoverer.getTargetModel();
...
\end{lstlisting}

\begin{figure*}[!h]
	\begin{center}
		\includegraphics[width=\textwidth]{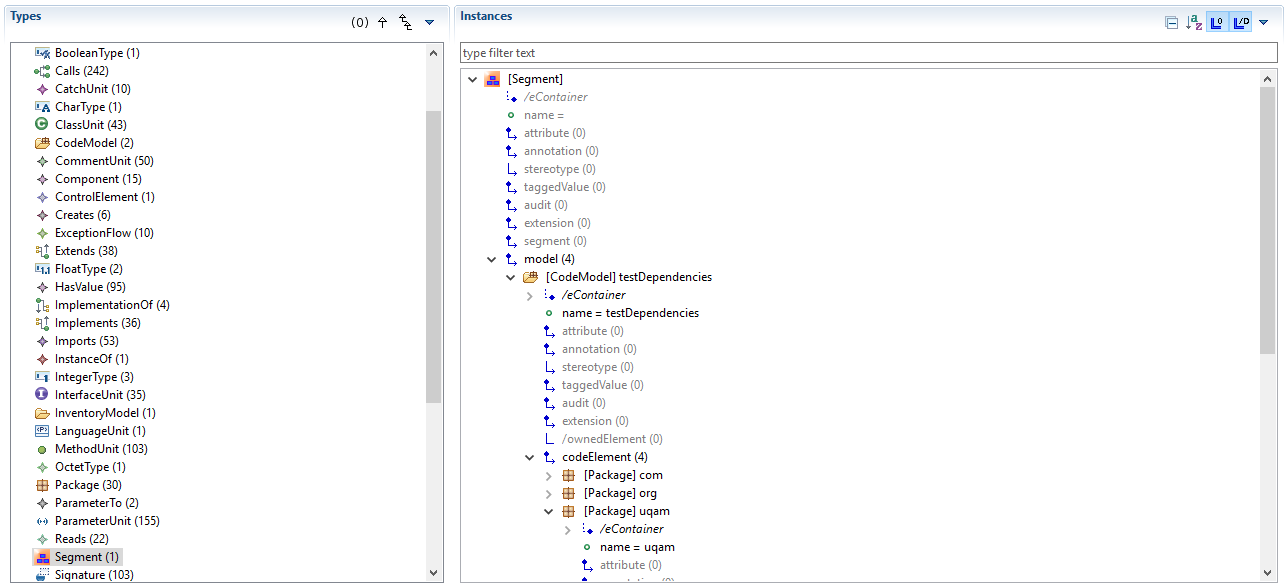}
		\caption{Example of a KDM Model extracted from Java source code}
		\label{fig:kdm-example}
	\end{center}
\end{figure*}

Figure \ref{fig:kdm-example} presents an example of a KDM model extracted based on the static analysis of the source code of a given Java project.

For more information about MoDisco, we refer to main the official website of MoDisco \cite{modisco} that contains the presentation and the documentation of MoDisco, and the MoDisco's developer forum \cite{modisco:developer:forum} that provides solutions of a bunch of problems that the users of MoDisco faced.

\section{The Codification of Dependencies Related to JSP Tags}
\label{sec:codification-remaining-tags}
There are a set of dependencies that are not recovered as a result of the previous step. Such dependencies are related to some tags that are embedded in string literals. 

In this section, we present (1) the set of tags related to dependencies in JSP pages, (2) the identification of related dependencies by parsing these tags, and (3) the representation of these identified dependencies in the KDM model identified based on the previous step (c.f. Section \ref{sec:reduction-problem}).

\subsection{The Set of Tags Related to Dependencies in JSP}
In \cite{shatnawi2017:tch:jeetechnology} \cite{shatnawi2017analyzing}, we identified the set of JSP tags that are used to make dependencies to other server pages\footnote{For more details about these tags, please refer to \cite{shatnawi2017:tch:jeetechnology}}. These are:

\begin{enumerate}

\item \textbf{The HTML form} (\textit{<form action="/myPage.jsp" method="get">}) that forwards requests to server pages. The URL of the target server page URL is specified in the \textit{\textbf{action}} attribute and the \textit{\textbf{method}} attribute defines the request type; \textit{get}, \textit{put} or \textit{post}.

\item \textbf{The \textit{jsp:include} action tag} (\textit{<jsp:include page="/myPage.jsp." flush="true" />}) that includes the content of server page of the relative-URL attached to the \textit{\textbf{page}} attribute.

\item \textbf{The include directive} tag (\textit{<\%@ include file="/myPage.jsp" \%>} or \textit{<jsp:directive.include file="/myPage.jsp" />}) that merges the content of other server pages. The \textit{\textbf{file}} attribute defines the related server page URL. 	

\item \textbf{The \textit{jsp:forward} action tag} (\textit{<jsp:forward page="myPage.jsp" />}) that forwards requests to other server pages. The target server page is determined based on a relative-URL attached to the \textit{\textbf{page}} attribute. 

\item \textbf{The \textit{page} directive tag} (\textit{<\%@ page errorPage="errorPage.jsp" \%>} or \textit{<jsp:directive.page errorPage="errorPage.jsp"/>}) that refer to a server page to be called in the case of exceptions. The \textit{\textbf{errorPage}} attribute defines the URL of this server page. 

\item \textbf{The \textit{href} tag} (\textit{<a href="https://www.uqam.ca">}) that makes a link to an URL.

\item \textbf{Two tags from the stander JSTL core tag library} that contains two interesting tags: the \textit{\textbf{<c:redirect url="relative-URL"/>}} and the \textit{\textbf{<c:url value="/myPage.jsp"}} tag.
	
\end{enumerate}

The set of tags and their interesting attributes are shown in Table \ref{table:tags-and-attributes}.

\begin{table}[h]
\centering
\caption{Summarization of the list of tags and their attributes related to dependencies}
\label{table:tags-and-attributes}
\begin{tabular}{|l|l|}
\hline
\multicolumn{1}{|c|}{\textbf{Tags}}                 & \multicolumn{1}{c|}{\textbf{Attributes}} \\ \hline
\textless form\textgreater                  & action, method                  \\ \hline
\textless jsp:include\textgreater           & page                            \\ \hline
\textless \%@ include\textgreater           & file                            \\ \hline
\textless jsp:directive.include\textgreater & file                            \\ \hline
\textless jsp:forward\textgreater           & page                            \\ \hline
\textless \%@ page \%\textgreater           & errorPage                       \\ \hline
\textless jsp:directive.page\textgreater    & errorPage                       \\ \hline
\textless a\textgreater                     & href                            \\ \hline
\textless c:redirect\textgreater            & url                             \\ \hline
\textless c:url\textgreater                 & value                           \\ \hline
\end{tabular}
\end{table}

\subsection{The Identification of Related Dependencies by Parsing Tags}
We identify the dependencies of each JSP page based on: (1) the extraction of a set of URLs invoked by the JSP page, and (2) the mapping of each URL to the corresponding server page. 

\subsubsection{The Extraction of a Set of URLs Invoked by Each JSP Page}
\begin{figure*}[!h]
	\begin{center}
		\includegraphics[width=\textwidth]{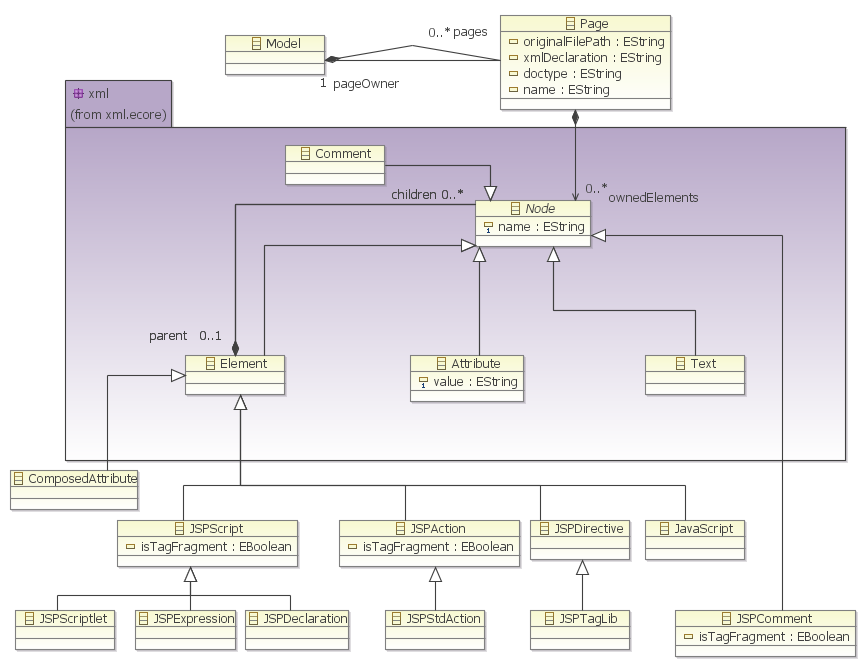}
        \caption{The metamodel of the JSP Model of MoDisco \cite{modisco}}
		\label{fig:jsp-metamodel}
	\end{center}
\end{figure*}

We develop a tool that parses JSP pages to identify a set of URLs of server pages that each JSP page depends on. 
Our tool relies on the \textit{JSP Model} offered by MoDisco. As it is shown in the JSP meta-model used by MoDisco and presented in Figure \ref{fig:jsp-metamodel}, the JSP model only represents tags of each JSP page in terms of XML tags.  However, it does not identify dependencies between JSP tags and across the JSP pages. In addition, it does not make dependencies with the other J2EE technologies including Beans components and other Java source code. 

Instead of dealing with textual JSP code, we use this JSP model as intermediate representation that facilitates the parsing process of JSP pages. Then, we propose an algorithm to travel through this JSP model to identify the set of URLs invoked by each JSP page based on the list presented in Table \ref{table:tags-and-attributes}. 

\begin{figure*}[!h]
	\begin{center}
		\includegraphics[width=\textwidth]{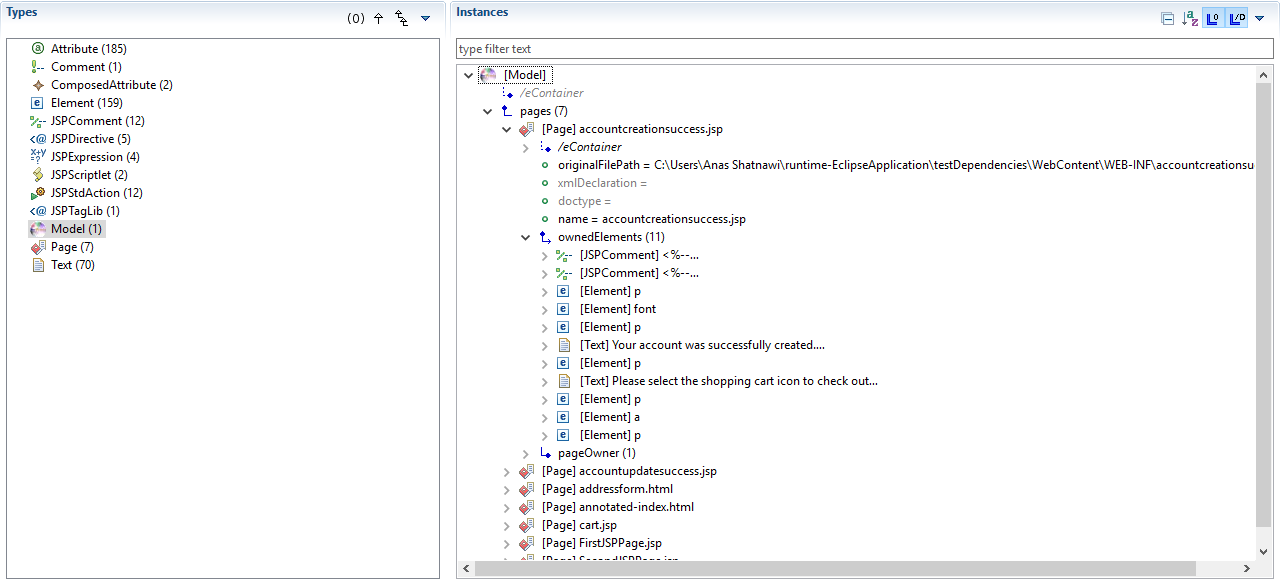}
		\caption{Example of a JSP Model extracted based on MoDisco}
		\label{fig:jsp-model}
	\end{center}
\end{figure*}

A JSP model can be programmatically identified based on the \textit{JSP Discoverer} presented in Listing \ref{JSP-discoverer-implementation}. An example of a JSP model extracted based on MoDisco is presented in Figure \ref{fig:jsp-model}.

\begin{lstlisting}[caption=Example of code that implements the MoDisco's JSP discoverer, label=JSP-discoverer-implementation]
...
DiscoverJspModelFromJavaElement discoverer = new DiscoverJspModelFromJavaElement();
discoverer.discoverElement(javaProject, monitor);
Resource jspResource = discoverer.getTargetModel();
...
\end{lstlisting}

\subsubsection{The Mapping of URLs to the Corresponding Server Pages}
\begin{figure*}[h]
	\begin{center}
		\includegraphics[width=\textwidth]{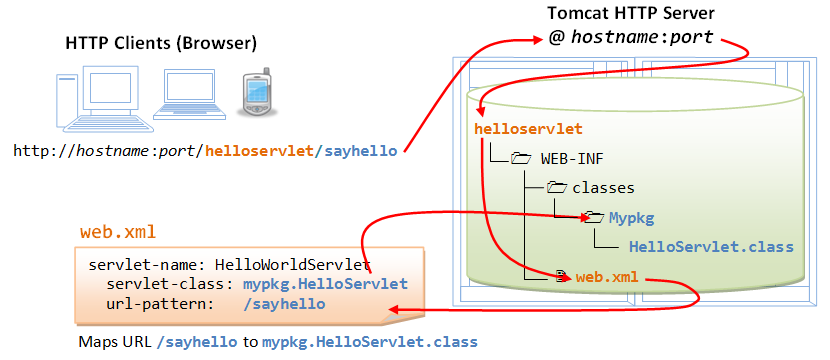}
		\caption{Mapping URLs to server pages}
		\label{fig:url-mapping}
	\end{center}
\end{figure*}

Figure \ref{fig:url-mapping} explains the process of mapping URLs to server pages. The URLs are mapped to server pages either using the \textit{web.xml}\footnote{The \textit{web.xml} is located in the \textit{WEB-INF/} directory of a JEE application} file (for Servlets, JSPs and JSFs) and/or \textit{@WebServlet} annotation\footnote{Starting from Servlet 3.0 specification, the \textit{@WebServlet} annotations are located within the Servlet's source code} (for Servelts) \cite{shatnawi2017:tch:jeetechnology}. Consequently, it is essential to visit (and parse) \textit{web.xml} and \textit{@WebServlet} annotations.

In \textit{web.xml}, we consider the five elements to identify mappings between server pages and relative-URLs, namely: \textit{<Servlet>},  \textit{<Servlet-mapping>}, \textit{<Servlet-class>}, \textit{<jsp-file>} and \textit{<url-pattern>}. Then, we developed a parser\footnote{Based on the SAX2 APIs: http://www.java2s.com/Code/Java/Servlets/

ParseawebxmlfileusingtheSAX2API.htm} to extract the needed from the \textit{web.xml} file.
Next, we built a look-up table that maps each URL to its corresponding server page(s).

\subsection{The Representation of Dependencies of JSP Tags in KDM Model}
In the previous steps, we identified the list of server pages that each JSP depends on. Now, we want to modify the resulting KDM model from the first step (c.f., Section \ref{sec:reduction-problem}) to add dependencies of server pages. 
We perform as follows: 

\begin{enumerate}
\item For each JSP page, we identify the KDM \textit{ClassUnit} instances that is related to the current JSP page and the list of target server pages in the KDM model\footnote{KDM \textit{ClassUnit} instances were generated by the translated Servlets}. We rely on a sequential search algorithm to identify these \textit{ClassUnit} instances. 

\item Then, in the \textit{service} method of the current JSP, we create new method invocations to the \textit{service()} methods of \textit{ClassUnit} instances of the corresponding server pages. Listing \ref{adding-mothed-invocation} shows an example of code that add a method invocation between two {ClassUnit} instances of two JSP pages.
\end{enumerate}

\begin{lstlisting}[caption=Example of code that add method invocation between \textit{ClassUnit} instances of two JSP pages, label=adding-mothed-invocation]
	public void addMethodCallBetweenClasses(ClassUnit caller, ClassUnit targetClass) {
		if  (caller == null){
			return;
		}
		EList<CodeItem> jspServletKdm = caller.getCodeElement();
		for (CodeItem cItem : jspServletKdm) {
			if (cItem instanceof MethodUnit) {
				// select _jspService() method
				if (cItem.getName().equals("_jspService")) {
					EList<AbstractCodeElement> mElements = ((MethodUnit) cItem).getCodeElement();
					for (AbstractCodeElement element : mElements) {
						if (element instanceof BlockUnit) {
							CodeRelationship call = CodeFactory.eINSTANCE.createCodeRelationship();
							call.setTo(targetClass);
							call.setFrom(caller);
							CodeElement actionElement = CodeFactory.eINSTANCE.createCodeElement();
							actionElement.setName("newCall");
							actionElement.getCodeRelation().add((AbstractCodeRelationship) call);
							((BlockUnit) element).getCodeElement().add(actionElement);
							element.getCodeRelation().add(call);
						}
					}
				}
			}
		}
	}
\end{lstlisting}

\section{Conclusion}
\label{sec:conclusion}
Existing tools do not identify KDM models of JSP pages. Thus, we proposed, in this report, an approach that identifies a KDM model of JSP pages. 

We first translated JSP pages to Java Servlets to reduce the problem space by mapping our problem to an already solved one. To do so, we defined \textit{ant} code based on the \textit{Jasper} tool provided by Apache Tomcat Server. Then, we developed a set of tools that:

\begin{itemize}
\item Identify the list of URLs invoked by each JSP page.

\item Extract the mapping of URLs to server pages based on parsing the \textit{web.xml} and \textit{@WebServlet}.

\item Read and modify the KDM model to include the JSP dependencies.

\end{itemize}

As a future direction, we want to develop an approach that identified KDM model for other server pages such as JSFs.

\bibliographystyle{unsrt}
\bibliography{main}  

\end{document}